\documentclass[11pt, a4paper]{article}
\usepackage[text={8in,11in},margin=1.0in,includefoot]{geometry}
\usepackage{doc}
\usepackage{url}
\usepackage{graphicx}
\usepackage{epstopdf}
\usepackage{amsmath}
\usepackage{pdflscape}
\usepackage{float}
\usepackage{placeins}
\restylefloat{table}

\numberwithin{equation}{section}
\title{Magnetic moments of $J^{P}=\frac{3}{2}^{+}$ decuplet baryons using statistical model}
\author{Amanpreet Kaur, Alka Upadhyay
\\\small{\it School of Physics and Material Science},\\\small{\it Thapar University,
Patiala, Punjab-147004}\\\small{E-mail:
amanpreet.kaur9074@yahoo.com, alka.iisc@gmail.com}}

\begin{document}
\date{}

\maketitle
\begin{abstract}A suitable wave function for baryon decuplet is framed with
inclusion of sea containing quark-gluon Fock states. Relevant
operator formalism is applied to calculate magnetic moments of
$J^{P}=\frac{3}{2}^{+}$ baryon decuplet. Statistical model assumes
decomposition of baryonic state in various quark-gluon Fock states
such as $|qqq\rangle|g\rangle, |qqq\rangle|gg\rangle,
|qqq\rangle|ggg\rangle$ with possibility gluon emitting
$q\overline{q}$ pairs condensates due to the subprocesses like
$g\Leftrightarrow q\overline{q}, g\Leftrightarrow gg$ and
$g\Leftrightarrow qg$ where $q\overline{q}=u\overline{u},
d\overline{d}, s\overline{s}$. Statistical approach and detailed
balance principle in combination is used to find the relative
probabilities of these Fock states in flavor, spin and color space.
The total number of partons (sea) in this formalism are restricted
to three gluons due to limited free energy of gluon and suppressed
number of strange quark-antiquark pairs. The combined approach is
used to calculate the magnetic moments, importance of strangeness in
the sea (scalar, vector and tensor). Our approach has confirmed the
scalar-tensor sea dominancy over vector sea. Various modifications
in the model are used to check the validity of statistical approach.
The results are matched with theoretical data available. Good
consistency with the experimental data have been achieved for
$\Delta^{++}$, $\Delta^{+}$ and $\Omega^{-}$.

\noindent
\vspace{0.2cm}
\end{abstract}
\section{Introduction and Motivation}
\label{introduction} Recently, a new state of matter called
"pentaquarks" $(uudc\overline{c})$ has been predicted at LHCb
bringing a revolution in the study of baryon spectroscopy. The
search from the LHCb was motivated by the prediction made by
theoretical approaches. Besides this the prediction of lifetime of
spin 3/2 heavy baryon state at CMS [1] helps us to explore their
properties in a better way. Lot of progress has been observed in
both theoretical and experimental approaches for the study of hadron
properties since octet magnetic moments were predicted by Coleman
and Glashow [2] about fifty years ago. These predictions motivated
theorist and experimentalist to measure baryon octet magnetic
moments [3]. The experimental information about decuplet baryons is
limited because they have short lifetimes so till now the
experimental data of $\Delta^{++}, \Delta^{+}, \Omega^{-}$ is
available so far [4-7]. The study about the properties of baryon
constitute an important role for investigation of baryon structure.
The advancements in the experimental facilities at CDF [8] etc. have
become a subject of motivation to study baryon properties and hence
its structure in the non perturbative regime of Quantum
Chromodynamics (QCD). The well known experiments like EMC
(Electron–Muon Collaboration) and the SMC (Spin-Muon
Collaboration)[9-10] studies the static properties of the hadrons.

The magnetic moments of baryon decuplet have been theoretically
investigated using different approaches, such as: simple additive
quark model in non relativistic limit which calculates the magnetic
moments of the baryons as the sum of its constituent quark magnetic
moments. Further improvements were done by including effects such as
sea quark contributions [11], quark orbital momentum effects [12],
SU(3) symmetry breaking effects [13]. Sogami and Oh'yamaguchi [14],
presented a concept of effective mass to calculate magnetic moments
of baryons and later, Bains and Verma [15] used the concept of
effective mass and screened charge of quarks to calculate magnetic
moments. The subject of magnetic moments is bit difficult to explain
or understand because this phenomenon of baryons is contributed from
the magnetic moments of valence quarks as well as from various other
complicated effects such as relativistic effects, contributions from
pion cloud, confinement effect on quark masses, etc..

Recently, predictions based on a number of theoretical formalisms
have been developed to calculate the magnetic moment of decuplet
baryons. The relativistic quark model (RQM) [16], QCD-based quark
model (QCDQM) [17], effective mass scheme (EMS) [18], light cone QCD
sum rule (LCQSR) [19], QCD sum rule (QCDSR) [20], Skyrme model [21],
chiral quark soliton model (CQSM) [22], chiral perturbation theory
($\chi$PT) [23], lattice QCD (LQCD) [24-25] etc. are few to name.
Discrete methods are suggested in literature [11] which describes
baryon to be composed of three valence quarks and sea composed of
gluons and quark antiquark pairs which have studied the nucleonic
properties like magnetic moments, weak decay coupling ratios, spin
distributions by considering the contributions from sea as well.

The strange quark to non strange quark ratio:
$\frac{2(s+\overline{s})}{(u+\overline{u}+d+\overline{d})}=0.477\pm0.063\pm0.053$[26]
predicted by the NuTeV Collaboration at FermiLab shows the presence
of strange quarks in sea. The contributions of strange quark to the
nucleonic form factors is negligible as suggested by the
collaborations HAPPEX and GO [27-29] and some theoretical studies as
well. Later R.Bijker et al. [30] suggested that contribution of
strange quark–antiquark pair is very less (-0.0004$\mu_{N}$) to
magnetic moment of the nucleon.

To calculate the magnetic moments of baryon decuplet particles , we
assume the baryon to be comprising of valence part and a virtual sea
consisting of quark antiquark pairs multiconnected by gluons. In
sec. II, a suitable wavefunction is framed for baryon decuplet
having color, flavor and spin space. Here, valence $q^{3}$ and a sea
combines in a way to reproduce desired quantum numbers of the
decuplets i.e. spin 3/2, color singlet and flavor 10. Sec. III shows
the application of magnetic moment operator to the total
wavefunction on flavor and spin space. Principle of detail balance
and statistical approach is applied in combination to find the
relative probabilities in spin, flavor and color states in sec. IV.
Here, the detailed balance is used to put a constraint
$(1-C_{l})^{n-1}$ on sea to be taking up $s\overline{s}$ pair due to
their heavy masses in terms of on respective baryons. Various
modifications in the statistical model is studied in sec.V. In sec.
VI, the calculated results have been analyzed and matched with the
theoretical models and experimental data.

\section{Decuplet wave function with a sea component}
The structure of hadron constitutes two parts i.e valence part (qqq)
and other is sea part which consist of quark-antiquark pairs
muticonnected through gluons (g,$q\overline{q}$) [31-32]. A $q^{3}$
state in the baryon are in the 1, 8 and 10 color states which means
that sea should also be in corresponding states to form a color
singlet baryon. The valence part of the hadronic wave function can
be written as:
\begin{equation}
    \label{eq:wavefunction}
  \Psi=\Phi(|\phi\rangle|\chi\rangle|\psi\rangle)(|\xi\rangle)
     \end{equation}where $|\phi\rangle,|\chi\rangle,|\psi\rangle$ and $|\xi\rangle$
denote flavor, spin, color and space $q^3$ wave functions and their
contribution make total wave function antisymmetric in nature. Here,
spatial part $(|\xi\rangle)$ is symmetric under the exchange of any
two quarks for the lowest lying hadrons and therefore the
flavor-spin-color part $\Phi(|\phi\rangle|\chi\rangle|\psi\rangle)$
should be antisymmetric in nature such that when combined with
$(|\xi\rangle)$ gives antisymmetry of the total wave function.

Sea considered here is in S-wave state with spin (0,1,2) and color
(1,8,$\overline{10}$) and is assumed to be flavorless . Let
$H_{0,1,2}$ and $G_{1,8,\overline{10}}$ denote spin and color sea
wave functions, which satisfy $\langle
H_{i}|H_{j}\rangle=\delta_{ij}$ , $\langle
G_{k}|G_{l}\rangle=\delta_{kl}$. In this approach we have assumed a
sea to be consist of two gluons or $q\overline{q}q\overline{q}$
pairs and different possible states for them can be written as:
\begin{gather*}
\textbf{Spin}: uud: 1/2\otimes1/2\otimes1/2=2(1/2)\oplus3/2\\
gg:1\otimes1=0_{s}\oplus1_{a}\oplus2_{s}\\
q\overline{q}q\overline{q}:(1/2\otimes1/2)\otimes(1/2\otimes1/2)=(0_{a}\oplus1_{s})\otimes(0_{a}\oplus1_{s})=2(0_{s})\oplus1_{s}\oplus2(1_{a})\oplus2_{s}
\end{gather*}

\begin{gather*}
\textbf{Color}: uud:
3\otimes3\otimes3=1_{a}\oplus8_{ms}\oplus8_{ma}\oplus10_{s}\\
gg:
8\otimes8=1_{s}\oplus8_{s}\oplus8_{a}\oplus10_{a}\oplus\overline{10}_{a}\oplus27_{s}\\
q\overline{q}q\overline{q}:(3\otimes\overline{3})\otimes(3\otimes\overline{3})=(1_{a}\oplus8_{s})\otimes(1_{a}\oplus8_{s})\\=2(1_{s})\oplus2(8_{s})\oplus2(8_{a})\oplus10_{s}\oplus\overline{10}_{s}\oplus27_{s}
\end{gather*}
Subscripts s and a denotes symmetry and antisymmetry on combining
the states. We have assumed in our model that gluon and
$q\overline{q}$ carry same quantum numbers. Total antisymmetry of
the baryon should be kept in mind while combining the valence and
sea part. In general, the symmetry property arise when (S+S),(A+A)
combines and antisymmetry comes into play when (S+A) combination is
formed.

So,the possible combinations of valence q$^3$ and sea wave functions
which can yield spin 3/2, flavor decuplet and color singlet state
thereby maintaining the anti symmetrization of the total baryonic
wave function are:
\begin{equation}
\label{eq:wavefunction}
 \Phi_{1}^{(3/2)}H_{0}G_{1} ,\Phi_{1}^{(3/2)}H_{1}G_{1} ,
\Phi_{1}^{(1/2)}H_{1}G_{8}  ,\Phi_{1}^{(3/2)}H_{2}G_{1},
\Phi_{8}^{(1/2)}H_{2}G_{8}
\end{equation}
The total flavor-spin-color wave function of a spin up baryon
decuplet consisting of three valence quarks and a sea component can
be written as:

\begin{equation}
\begin{split} \label{eq:wavefunction}
|\Phi_{3/2}^{(\uparrow)}\rangle=\frac{1}{N}[a_{0}\Phi_{1}^{(3/2\uparrow)}H_{0}G_{1}+
b_{1}(\Phi_{1}^{(3/2)}\otimes H_{1})^{\uparrow}G_{1}+
b_{8}(\Phi_{1}^{(1/2)}\otimes H_{1})^{\uparrow}G_{8}+\\
d_{1}(\Phi_{1}^{(3/2)}\otimes H_{2})^{\uparrow}G_{1}+
d_{8}(\Phi_{8}^{(1/2)}\otimes H_{2})^{\uparrow}G_{8}]
\end{split}
\end{equation}

\begin{equation}
\label{eq:wavefunction}N^{2}=a_{0}^{2}+b_{1}^{2}+b_{8}^{2}+d_{1}^{2}+d_{8}^{2}
\end{equation}where N is the normalization constant.
The first term in the equation (2.3) is obtained by combining
$q^{3}$ wave function with spin 0 (scalar sea) and next two terms
are obtained by coupling $q^{3}$ with spin 1(vector sea) such that:
\begin{equation}
\label{eq:wavefunction} (\Phi_{1}^{(3/2)}\otimes
H_{1})^{\uparrow}\equiv\phi_{b1}^{(3/2\uparrow)}\psi_{1}^{A},
\end{equation}
\begin{equation}
\label{eq:wavefunction} (\Phi_{1}^{(1/2)}\otimes
H_{1})^{\uparrow}\equiv\phi_{b8}^{(1/2\uparrow)}\psi_{1}^{MS},
\end{equation}
where
\begin{equation}
\label{eq:wavefunction}
\phi_{b1}^{(3/2\uparrow)}=\sqrt{\frac{3}{5}}H_{1,0}F_{S}^{(3/2\uparrow)}-\sqrt{\frac{2}{5}}H_{1,1}F_{S}^{(1/2\uparrow)}
\end{equation}
\begin{equation}
\label{eq:wavefunction}
\phi_{b8}^{(1/2\uparrow)}=\sqrt{\frac{3}{5}}H_{1,0}F_{S}^{(3/2\uparrow)}-\sqrt{\frac{3}{5}}H_{1,1}F_{S}^{(1/2\uparrow)}
\end{equation}
The $q^{3}$ wave functions in equation (2.5-2.6) for a flavor
decuplet baryon can be written as:
\begin{equation}
\label{eq:wavefunction}
\Phi_{1}^{(3/2)}\equiv\Phi(10,3/2,1)=F_{S}\Psi_{1}^{A}
\end{equation}
where
\begin{equation} \label{eq:wavefunction}
F_{S}=\phi^{\lambda}\chi^{\lambda}
 \end{equation}
and
\begin{equation}
\label{eq:wavefunction}
\Phi_{1}^{(1/2)}\equiv\Phi(10,1/2,1)=F_{MS}\Psi_{1}^{A}
\end{equation}

\begin{equation}
\label{eq:wavefunction}
\Phi_{8}^{(1/2)}\equiv\Phi(10,1/2,8)=F_{A}\phi_{8}^{S}
\end{equation}
where
\begin{equation}
\label{eq:wavefunction}
  F_{MS}=\phi^{\lambda}\chi^{MS}
\end{equation}

\begin{equation}
\label{eq:wavefunction}
F_{A}=\frac{1}{\sqrt{2}}(\chi^{\lambda}\psi^{\rho}-\chi^{\rho}\psi^{\lambda})
\end{equation}
Here, superscripts S and A denote total symmetry and antisymmetry
and $\lambda$ , MS denotes symmetry and mixed symmetry under quark
permutations $q_{1}\leftrightarrow q_{2}$. $\Phi_{1}^{3/2}$ denotes
a function with spin 3/2, color singlet and 10 represents flavor
part. This function can be written as combination of $F_{s}$
(denotes flavor and spin) and $\Psi_{1}^{A}$ (represents color of
baryons and is antisymmetric). For $F_{s}$ to be symmetric, $\phi$
and $\chi$ should be symmetric in nature. Similarly, other functions
like $\Phi_{1}^{1/2}$ and $\Phi_{8}^{1/2}$ denotes a function with
spin 1/2 and flavor 10 with color singlet and octet respectively.
Each wave function is a combination of symmetric and antisymmetric
term such that total wave function becomes antisymmetric in nature.

The final two terms are result of coupling with spin 2 (tensor sea).
Their expressions can be written as:
\begin{equation}
\label{eq:wavefunction} (\Phi_{1}^{(3/2)}\otimes
H_{2})^{\uparrow}\equiv\phi_{d1}^{(3/2\uparrow)}\psi_{1}^{A},
\end{equation}
\begin{equation}
\label{eq:wavefunction} (\Phi_{8}^{(1/2)}\otimes
H_{2})^{\uparrow}\equiv\phi_{d8}^{(1/2\uparrow)}\phi_{8}^{S},
\end{equation}
where
\begin{equation}
\label{eq:wavefunction}
\phi_{d1}^{(3/2\uparrow)}=\sqrt{\frac{1}{5}}H_{2,0}F_{S}^{(3/2\uparrow)}-\sqrt{\frac{2}{5}}H_{2,1}F_{S}^{(1/2\uparrow)}+\sqrt{\frac{2}{5}}H_{2,2}F_{S}^{(1/2\downarrow)}
\end{equation}
\begin{equation}
\label{eq:wavefunction}
\phi_{d8}^{(1/2\uparrow)}=\sqrt{\frac{1}{5}}H_{2,0}F_{A}^{(3/2\uparrow)}-\sqrt{\frac{2}{5}}H_{2,1}F_{A}^{(1/2\uparrow)}+\sqrt{\frac{2}{5}}H_{2,2}F_{A}^{(1/2\downarrow)}
\end{equation}
Wave functions
$\phi_{b1}^{(3/2\uparrow)}$,$\phi_{b8}^{(1/2\uparrow)}$,$\phi_{d1}^{(3/2\uparrow)}$,$\phi_{d8}^{(1/2\uparrow)}$
are the functions written by taking coupling between spin of sea
part and flavor part of $q^{3}$ wave function. The coefficients
associated with each term contains the information about magnetic
moments, spin distribution among valence quarks and needs to be
determined statistically. The parameter $a_{0}$ come from a spin 3/2
of $q^{3}$ state coupled to spin 0 (scalar) of sea, $b_{1}$,$b_{8}$
comes when spin 3/2 and 1/2 of $q^{3}$ state is coupled to spin 1
(vector)of sea and $d_{1}$,$d_{8}$ corresponds coupling of spin 3/2
to spin 2 (tensor) of sea. The idea of different coefficients for
each baryon wave function of the baryon decuplet comes from the fact
that each baryon have different mass and quark content.
\section{Magnetic Moments}
Magnetic moments is a property of hadrons observed at low energies
and long distances. Magnetic moments are contributed by all
constituents of baryon (valence+sea) contributes by experiencing the
same magnetic field. Thus, for baryons at ground state, the magnetic
moments is a vector sum of quark magnetic moments,
\begin{equation}
\label{eq:wavefunction} \mu_{baryon}=\sum_{i=1,2,3}\mu_{i}\sigma_{i}
\end{equation}
where $\sigma_{i}$ is the pauli matrix representing the spin term of
$i^{th}$ quark and $\mu_{i}$ represents magnitude of quark magnetic
moments and therefore values of magnetic moments are different for
different baryons. Also,
\begin{equation}
\label{eq:wavefunction} \mu_{baryon}=\mu_{i}=\frac{e_{i}}{2m_{i}}
\end{equation}
for i = u,d,s and $e_{i}$ represents the quark charge. Present work
shows the calculation of magnetic moments of $J^{P}=\frac{3}{2}^{+}$
by applying the magnetic moment operator
$(\widehat{O}=\mu_{i}\sigma_{i})$ to the baryon wave function in a
way:
\begin{equation}
\label{eq:wavefunction}
\begin{split}
\langle\Phi_{3/2}^{\uparrow}|\widehat{O}|\Phi_{3/2}^{\uparrow}\rangle=\frac{1}{N^{2}}[a_{0}^{2}\Phi_{1}^{(3/2\uparrow)}|\widehat{O}|\Phi_{1}^{(3/2\uparrow)}\rangle+b_{1}^{2}\Phi_{1}^{(3/2\uparrow)}|\widehat{O}|\Phi_{1}^{(3/2\uparrow)}\rangle\\+b_{8}^{2}\Phi_{1}^{(1/2\uparrow)}|\widehat{O}|\Phi_{1}^{(1/2\uparrow)}\rangle+d_{1}^{2}\Phi_{1}^{(3/2\uparrow)}|\widehat{O}|\Phi_{1}^{(3/2\uparrow)}\rangle+d_{8}^{2}\Phi_{8}^{(1/2\uparrow)}|\widehat{O}|\Phi_{8}^{(1/2\uparrow)}\rangle]
\end{split}
\end{equation}
The magnetic moment operator $(\widehat{O}$) which depends on flavor
and spin of $i^{th}$ quark can be applied on the baryon wave
function in a following way:
\begin{equation}
\begin{split}
\label{eq:wavefunction}
\langle\Phi_{3/2}^{\uparrow}|\widehat{O}|\Phi_{3/2}^{\uparrow}\rangle=\frac{1}{N^{2}}[a_{0}\langle
O_{f}^{i}\rangle^{\lambda\lambda}\langle\sigma_{Z}^{i}\rangle^{\lambda\uparrow\lambda\uparrow}+b_{1}\langle
O_{f}^{i}\rangle^{\lambda\lambda}\langle\sigma_{Z}^{i}\rangle^{\lambda\uparrow\lambda\uparrow}+b_{8}\langle
O_{f}^{i}\rangle^{\lambda\lambda}\langle\sigma_{Z}^{i}\rangle^{\lambda\uparrow\lambda\uparrow}\\+d_{1}\langle
O_{f}^{i}\rangle^{\lambda\lambda}\langle\sigma_{Z}^{i}\rangle^{\lambda\uparrow\lambda\uparrow}+d_{8}\langle
O_{f}^{i}\rangle^{\lambda\lambda}\langle\sigma_{Z}^{i}\rangle^{\lambda\uparrow\lambda\uparrow}]
\end{split}
\end{equation}

\setlength{\tabcolsep}{0.5em} %
{\renewcommand{\arraystretch}{1.5}%
\FloatBarrier
\begin{table}[H]

\caption{Expressions obtained after applying magnetic moment
operator to baryon decuplet are shown:}
\begin{center}
\begin{tabular}{|c|p{14cm}|}
\hline Baryon & $\frac{
\langle\Phi_{3/2}^{\uparrow}|\widehat{O}|\Phi_{3/2}^{\uparrow}\rangle
}{N^{2}}$\\ \hline \hline
  $\Delta^{++}$ & $a_{0}^{2}(15\mu_{u})+b_{1}^{2}(11\mu_{u})+b_{8}^{2}(11\mu_{u})+d_{1}^{2}(3\mu_{u})+d_{8}^{2}(\frac{3}{2}\mu_{u})$ \\
  $\Delta^{+}$&  $a_{0}^{2}(30\mu_{u}+15\mu_{d})+b_{1}^{2}(22\mu_{u}+11\mu_{d})+b_{8}^{2}(22\mu_{u}+11\mu_{d})+d_{1}^{2}(8\mu_{u}+\mu_{d})+d_{8}^{2}(4\mu_{u}+\frac{1}{2}\mu_{d})$\\
  $\Delta^{0}$ &  $a_{0}^{2}(30\mu_{d}+15\mu_{u})+b_{1}^{2}(22\mu_{d}+11\mu_{u})+b_{8}^{2}(22\mu_{d}+11\mu_{u})+d_{1}^{2}(8\mu_{d}+\mu_{u})+d_{8}^{2}(4\mu_{d}+\frac{1}{2}\mu_{u})$\\
  $\Delta^{-}$ & $a_{0}^{2}(15\mu_{d})+b_{1}^{2}(11\mu_{d})+b_{8}^{2}(11\mu_{d})+d_{1}^{2}(3\mu_{d})+d_{8}^{2}(\frac{3}{2}\mu_{d})$ \\
  $\Sigma^{*+}$ & $a_{0}^{2}(30\mu_{u}+15\mu_{s})+b_{1}^{2}(22\mu_{u}+11\mu_{s})+b_{8}^{2}(22\mu_{u}+11\mu_{s})+d_{1}^{2}(8\mu_{u}+\mu_{s})+d_{8}^{2}(4\mu_{u}+\frac{1}{2}\mu_{s})$ \\
  $\Sigma^{*0}$ & $a_{0}^{2}[5(\mu_{u}+\mu_{d}+\mu_{s})]+b_{1}^{2}[\frac{11}{3}(\mu_{u}+\mu_{d}+\mu_{s})]+b_{8}^{2}[\frac{11}{3}(\mu_{u}+\mu_{d}+\mu_{s})]+d_{1}^{2}(\mu_{u}+\mu_{d}+\mu_{s})+d_{8}^{2}[\frac{1}{2}(\mu_{u}+\mu_{d}+\mu_{s})]$ \\
  $\Sigma^{*-}$ & $a_{0}^{2}(30\mu_{d}+15\mu_{s})+b_{1}^{2}(22\mu_{d}+11\mu_{s})+b_{8}^{2}(22\mu_{d}+11\mu_{s})+d_{1}^{2}(8\mu_{d}+\mu_{s})+d_{8}^{2}(4\mu_{d}+\frac{1}{2}\mu_{s})$\\
  $\Xi^{*0}$ & $ a_{0}^{2}(30\mu_{s}+15\mu_{u})+b_{1}^{2}(22\mu_{s}+11\mu_{u})+b_{8}^{2}(22\mu_{s}+11\mu_{u})+d_{1}^{2}(8\mu_{s}+\mu_{u})+d_{8}^{2}(4\mu_{s}+\frac{1}{2}\mu_{u})$ \\
  $\Xi^{*-}$&  $ a_{0}^{2}(30\mu_{s}+15\mu_{d})+b_{1}^{2}(22\mu_{s}+11\mu_{d})+b_{8}^{2}(22\mu_{s}+11\mu_{d})+d_{1}^{2}(8\mu_{s}+\mu_{d})+d_{8}^{2}(4\mu_{s}+\frac{1}{2}\mu_{d})$\\
  $\Omega^{-}$ & $a_{0}^{2}(15\mu_{s})+b_{1}^{2}(11\mu_{s})+b_{8}^{2}(11\mu_{s})+d_{1}^{2}(3\mu_{s})+d_{8}^{2}(\frac{3}{2}\mu_{s})$ \\\hline
\end{tabular}
\end{center}
\end{table}
\FloatBarrier
where, $O_{f}^{i}=\frac{e^{i}}{2m_{i}}$ for magnetic
moments. The values of quark magnetic moments predicted by quark
model used in our approach are :\begin{equation}
\label{eq:wavefunction} \mu_{u}=1.852\mu_{N}, \mu_{d}=-0.972\mu_{N},
\mu_{s}=-0.613\mu_{N}
\end{equation}

\section{Principle of detailed balance and Statistical Model}
Principle of detailed balance proposed by Zhang et al. [33-34]
calculates the probability of Fock states present inside hadrons.
Detailed balance principle demands equality between arriving in from
one substate and leaving it. Hadron is treated to be consisting of
complete set of quark gluon Fock states and can be expressed in
expanded form as:
\begin{equation}
\label{eq:wavefunction}
|B\rangle=\sum_{i,j,k}C_{i,j,k}|(q),(i,j,k),(l)\rangle
\end{equation}
where {q} represents the valence quarks of the baryon , i is the
number of quark-antiquark $u\overline{u}$ pairs, j is the number of
quark-antiquark $d\overline{d}$ pairs , k is the number of gluons
and {l} is the number of $s\overline{s}$ pairs in sea. The
probability to find a quark-gluon Fock states is:
\begin{equation}
\label{eq:wavefunction} \rho_{i,j,k,l}=|C_{i,j,k,l}|^{2},
\end{equation} and
$\rho_{i,j,k,l}$ satisfies the normalization condition,
\begin{equation}
\label{eq:wavefunction} \sum_{i,j,k}\rho_{i,j,k,l}=1
\end{equation} Assumption of detailed balance principle is that every two subensembles balance with each other in a way:
\begin{equation}
\label{eq:wavefunction}
\rho_{i,j,l,k}|(q),(i,j,l,k)\rangle\overset{balance}
\rightleftharpoons\rho_{i^{'},j^{'},l^{'},k^{'}}|(q),(i^{'},j^{'},l^{'},k^{'}\rangle
\end{equation}The transfer between two Fock states has two ways: go-out rate and
come in rate which is proportional to number of partons that may
split and number of partons recombining respectively. The
calculation of probability distributions includes various sub
processes like $g\Leftrightarrow q\overline{q}$,$g\Leftrightarrow
gg$,$q\Leftrightarrow qg$. Detailed balance principle is applied to
$\Sigma^{*0}$ to calculate probabilities and can be written as :
\begin{enumerate}
    \item When $q\Leftrightarrow qg$ is considered: The
general expression of probability can be written as:
\begin{equation}
\label{eq:wavefunction}|uds,{i,j,l,k-1}\rangle
\underset{(3+2i+2j+2l)k}{\stackrel{3+2i+2j+2l}{\rightleftharpoons}}
|uds,{i,j,l,k}\rangle
\end{equation}
\begin{equation}
\label{eq:wavefunction}
\frac{\rho_{i,j,l,k}}{\rho_{i,j,l,k-1}}=\frac{1}{k}
\end{equation}
Here $\frac{\rho_{i,j,l,k}}{\rho_{i,j,l,k-1}}$ represents the
probability ratios of two processes.
\item When both the processes $g\Leftrightarrow gg$ and
$q\Leftrightarrow qg$ are included:
\begin{equation}
\label{eq:wavefunction}
|uds,{i,j,l,k-1}\rangle\underset{(3+2i+2j+2l)k+\frac{k(k-1)}{2}}{\stackrel{3+2i+2j+2l+k-1}{\rightleftharpoons}}
|uds,{i,j,l,k}\rangle
\end{equation}
\begin{gather}
\frac{\rho_{i,j,l,k}}{\rho_{i,j,l,k-1}}=\frac{3+2i+2j+2l+k-1}{3+2i+2j+2l)k+\frac{k(k-1)}{2}}
\end{gather}
\item When $g\Leftrightarrow q\overline{q}$ is considered: The
    processes $g\Leftrightarrow u\overline{u}$, $g\Leftrightarrow
    d\overline{d}$, $g\Leftrightarrow s\overline{s}$ are involved
    here.
\begin{gather}
\frac{\rho_{i,j,l,k}}{\rho_{i,j,l+k,0}}=\frac{(k(k-1)(k-2)......1(1-C_{0})^{n-2l-1}
(1-C_{1})^{n-2l}....(1-C_{l-1})^{n+k-2}}{(l+1)(l+2)....(l+k)(l+k+1)}
\end{gather}
\end{enumerate}
The details of the above calculations can be found in ref. [35]. The
subprocess $g\Leftrightarrow s\overline{s}$ is active only when they
satisfy the condition that gluons should have energy larger than at
least two times the mass of strange quark because strange quark has
non-negligible mass for gluons to undergo the process. The
subprocess $g\Leftrightarrow s\overline{s}$ is restricted by
applying the constraint defined as $k(1-C_{l})^{n-1}$ [34] which is
introduced from gluon free energy distribution. Here, n is the
number of partons present in the Fock state i.e n=3+2i+2j+l+2k in
our calculations. So taking $(1-C_{l})^{n-1}$ as the suppressing
factor for generating $s\overline{s}$ pairs from a gluon and by
using the detailed balance model the strange quark contribution to
the baryon can be calculated. The value of number of $s\overline{s}$
pairs has been restricted to two due to its large mass and limited
free energy of gluon undergoing the subprocess $q\Leftrightarrow s
\overline{s}$. Detailed balance principle when applied to different
baryons gives different results because sea content will split and
recombine with quark content which is different for every baryon.

The expressions of probabilities in terms of $\rho_{0,0,0,0}$ for
$\Sigma^{*0}$:
\begin{gather}
\frac{\rho_{i,j,l+k,0}}{\rho_{0,0,0,0}}=\frac{1}{i!(i+1)!j!(j+1)!(l+k)!(l+k+1)!}
\end{gather}
Similar expressions of probabilities for other decuplet particles
can be written in terms of $\rho_{0,0,0,0}$ and are shown below in
table 2.
\setlength{\tabcolsep}{0.5em} %
{\renewcommand{\arraystretch}{1.3}%
\begin{table}
\caption{Expressions for probabilities in terms of $\rho_{0,0,0,0}$
for $J^{P}=\frac{3}{2}^{+}$ decuplet:}
\begin{center}
\begin{tabular}{|c|p{4cm}|}
  \hline
  Baryon & $\frac{\rho_{i,j,l+k,0}}{\rho_{0,0,0,0}}$\\\hline
  $\Delta^{++}$ & $\frac{3}{i!(i+3)!j!j!(l+k)!(l+k)!}$ \\
  $\Delta^{+}$&  $\frac{2}{i!(i+2)!j!(j+1)!(l+k)!(l+k)!}$\\
  $\Delta^{0}$ &  $\frac{2}{i!(i+1)!j!(j+2)!(l+k)!(l+k)!}$\\
  $\Delta^{-}$ & $\frac{3}{i!i!j!(j+3)!(l+k)!(l+k)!}$ \\
  $\Sigma^{*+}$ & $\frac{2}{i!(i+1)!j!(j+1)!(l+k)!(l+k+1)!}$ \\
  $\Sigma^{*0}$ & $\frac{1}{i!(i+1)!j!(j+1)!(l+k)!(l+k+1)!}$ \\
  $\Sigma^{*-}$ & $\frac{2}{i!i!j!(j+2)!(l+k)!(l+k+1)!}$\\
  $\Xi^{*0}$ & $ \frac{2}{i!(i+1)!j!j!(l+k)!(l+k+2)!}$ \\
  $\Xi^{*-}$&  $ \frac{2}{i!i!j!(j+1)!(l+k)!(l+k+2)!}$\\
  $\Omega^{-}$ & $\frac{3}{i!i!j!j!(l+k)!(l+k+3)!}$ \\\hline
\end{tabular}
\end{center}
\end{table}The normalization condition $\sum_{i,j,k,l}\rho_{i,j,k,l}=1$ gives
the individual probabilities of baryon decuplets. The entire list of
probabilities of various Fock states i.e $\rho_{i,j,k,l}$ 's are
shown in table 3 for other decuplet members as well.

\setlength{\tabcolsep}{0.9em} %
{\renewcommand{\arraystretch}{1.3}%
\FloatBarrier
\begin{table}[H]
\caption{The values of probabilities of all Fock states i.e
$\rho_{i,j,k,l}$ 's. }
\begin{center}
\begin{tabular}{c|c|c|c|c|c|c}
  \hline
  i & j & L & $\rho_{i,j,k,l}$ & k=0 & k=1 & k=2 \\
  \hline
  0 & 0 & 0 & uds & 0.15454  &0.0992203  &0.00599579  \\\hline
  0&0  &1  &uds $s\overline{s}$  & 0.07727 & 0.0324754 & 0.00910211 \\\hline
  0&1  &0  &uds $d\overline{d}$  & 0.07727 & 0.0369213 &0.00157739  \\\hline
  1&0  &0  &uds $u\overline{u}$  & 0.07727 &0.0369213  &0.00157739  \\\hline
  0&1  &1  &uds $d\overline{d}s\overline{s}$  & 0.038635 & 0.01148 &0.00321758  \\\hline
  1&0  &1  &uds $u\overline{u}s\overline{s}$  &0.038635  &0.01148  &0.00321758  \\\hline
  1&1  &0  &uds $u\overline{u}d\overline{d}$  &0.038635  & 0.013739 & 0.000414987 \\\hline
  1&1  &1  &uds $u\overline{u}d\overline{d}s\overline{s}$  &0.0193175  &0.00405816  &0.00113741  \\\hline
  1&2  &0  &uds $u\overline{u}d\overline{d}d\overline{d}$  &0.00643917  &0.00170415  &0.0000363921  \\\hline
  0&2  &0  &uds $d\overline{d}d\overline{d}$  &0.0128783  &0.00457965  &0.000138329  \\\hline
  0&2  &1  &uds $d\overline{d}d\overline{d}s\overline{s}$  & 0.00643917 & 0.00135272 &0.000379137  \\\hline
  2&0  &0  &uds $u\overline{u}u\overline{u}$  & 0.0128783 &0.00457965  &0.000138329  \\\hline
  2&1  &0  &uds $u\overline{u}d\overline{d}$  &0.00643917  &0.00170415  & 0.0000363921 \\\hline
  2&0  &1  &uds $u\overline{u}u\overline{u}s\overline{s}$  & 0.00643917 &0.00135272 & 0.000379137 \\\hline
\end{tabular}
\end{center}
\end{table}
\FloatBarrier From the table 3, it is well observed that lesser
contributions from higher mass Fock states have suppressed the whole
Fock states with higher mass and SU(2) symmetry is well obeyed by
Fock states with u and d quarks. The number of $s\overline{s}$ pairs
have been limited to two because of heavy strange quark mass and
limited free energy of gluon as strange-antistrange pairs are
generated from subprocess $g\Leftrightarrow s\overline{s}$.

\medskip {\bf Statistical model} [36] is used in our formalism to calculate
magnetic moments of decuplet members by assuming hadrons as an
ensemble of quark gluon Fock states. We statistically decompose
quark-gluon Fock states $|q^{3},i,j,k,l\rangle$ of a baryon in a set
of states in which the valence part and sea have definite spin and
color quantum number. Statistical model has earlier been applied to
proton [36], assumes the baryon comprises of valence quarks plus a
virtual sea consisting of strange/non-strange quark antiquark pairs
multiconnected through gluons in the form of subprocesses like
$g\Leftrightarrow q\overline{q}$,$g\Leftrightarrow
gg$,$q\Leftrightarrow qg$.

The wave function in equation (2.3) can also be written in the form
of $\Phi_{val}\Phi_{sea}$ and the unknown parameters
$(a_{0},b_{1},b_{8},d_{1},d_{8})$ by a factor $\sum
n_{\mu\nu}^{*}c_{sea}$ such that total wave function becomes
$|\Phi_{\frac{3}{2}}^{\uparrow}\rangle=
\sum_{\mu,\nu}(n_{\mu\nu}^{*}c_{sea})\Phi_{val}\Phi_{sea}$ where
$\mu$ and $\nu$ have values 0, 1, 2 and 1, 8, $\overline{10}$
respectively. All $n_{\mu,\nu}'s$ are calculated from multiplicities
of each Fock state in spin and color space. These multiplicities are
expressed in the form of $\rho_{p,q}$ where relative probability for
core part should have angular momentum p and sea to have angular
momentum q such that the resultant angular momentum should come out
as 3/2. Similar probabilities could be calculated for color space
which yields color singlet state. Calculation of these probabilities
are expressed in the form of common factor ("c") for every
combination of valence and sea which is multiplied with multiplicity
factor (n) for each Fock state. The common parameter "c" can be
calculated from the table of various Fock states derived from the
principle of detailed balance. Each unknown parameter in the
equation of wave function will have a particular value of $\sum
n_{\mu\nu}c_{sea}$ depending on the Fock state: [37]
\begin{eqnarray*}
a_{0}=(n_{01}c_{sea})_{|gg\rangle}+(n_{01}c_{sea})_{|u\overline{u}g\rangle}+(n_{01}c_{sea})_{|d\overline{d}g\rangle}+(n_{01}c_{sea})_{|s\overline{s}g\rangle}+...\\b_{1}=(n_{11}c_{sea})_{|gg\rangle}+(n_{11}c_{sea})_{|u\overline{u}g\rangle}+(n_{11}c_{sea})_{|d\overline{d}g\rangle}+(n_{11}c_{sea})_{|s\overline{s}g\rangle}+...\\d_{1}=(n_{21}c_{sea})_{|gg\rangle}+(n_{21}c_{sea})_{|u\overline{u}g\rangle}+(n_{21}c_{sea})_{|d\overline{d}g\rangle}+(n_{21}c_{sea})_{|s\overline{s}g\rangle}+...\\............................................................................................................
\end{eqnarray*}
Combinations for other unknown parameters can be written in a
similar way. Though the set of different Fock states
($|gg\rangle,|u\overline{u}g\rangle,|d\overline{d}g\rangle$) etc. is
same for all baryon decuplet members but the probability
distribution is different for different baryons due to mass
inherited from flavor leading to different values of unknown
parameters. These calculations will give the value of a factor "nc"
for Fock states which has a significant role in determining the
magnetic moments. The subscripts s and a denotes symmetry and
anti-symmetry conditions. To find ratio of probabilities in spin and
color space, various decompositions are carried out in a way:

1. Consider the decomposition of state $|q^{3},0,0,0,2\rangle$ or
$|gg\rangle$ sea: Different cases of probability ratios for spin of
valence and sea can be written as:
\begin{eqnarray*}
\frac{\rho_{\frac{1}{2}, 1}}{\rho_{\frac{1}{2},
1}}=\frac{(\frac{3}{8}).(\frac{3}{9}).(\frac{1}{6})}{(\frac{3}{8}).(\frac{5}{9}).(\frac{2}{10})}=\frac{1}{2}\\
\frac{\rho_{\frac{3}{2}, 0}}{\rho_{\frac{3}{2},
1}}=\frac{(\frac{1}{8}).(\frac{1}{9}).(\frac{1}{4})}{(\frac{1}{8}).(\frac{3}{9}).(\frac{1}{8})}=2\\
\frac{\rho_{\frac{3}{2}, 0}}{\rho_{\frac{3}{2},
2}}=\frac{(\frac{1}{8}).(\frac{1}{9}).(\frac{1}{4})}{(\frac{1}{8}).(\frac{5}{9}).(\frac{1}{10})}=\frac{1}{2}
\end{eqnarray*}where $\rho_{p,q}$ is the probability that the
$q^{3}$ valence and gg sea are in spin states p and q respectively
which combine to give total spin as 3/2. The first term on the r.h.s
in the numerator or denominator denotes the relative probability for
valence quarks to have spin p and second term is for spin of sea (q)
and finally, third term denotes the resultant spin 3/2. Similar
probability ratios can be calculated for color spaces finally giving
a color singlet baryon and can be written as:
\begin{eqnarray*}
\frac{\rho_{1,1}}{\rho_{8,8_{s}}}=\frac{(\frac{1}{27}).(\frac{1}{64}).(1)}{(\frac{16}{27}).(\frac{8}{64}).(\frac{1}{64})}=\frac{1}{2}=\frac{\rho_{1,1}}{\rho_{8,8_{a}}}\\
\frac{\rho_{1,1}}{\rho_{10,\overline{10}}}=\frac{(\frac{1}{27}).(\frac{1}{64}).(1)}{(\frac{10}{27}).(\frac{10}{64}).(\frac{1}{100})}=1
\end{eqnarray*}
So, these are the probability distributions to find the valence
quarks in spin 3/2 and color singlet states with sea. To compute the
common parameter "c" the product of probabilities in spin and color
spaces can be written in terms of common factor "c" as;
\begin{eqnarray*}
\rho_{\frac{1}{2},1}[\rho_{8,8_{a}},\rho_{10,\overline{10}}]=c(2,1)\\
\rho_{\frac{1}{2},2}[\rho_{1,1},\rho_{8,8_{s}}]=2c(1,2)\\
\rho_{\frac{3}{2},0}[\rho_{1,1},\rho_{8,8_{s}}]=2c(1,2)\\
\rho_{\frac{3}{2},1}[\rho_{8,8_{a}}]=2c\\
\rho_{\frac{3}{2},2}[\rho_{1,1},\rho_{8,8_{s}}]=2c(1,2)
\end{eqnarray*}
Values present on the r.h.s of the above equations are the
multiplicities for a particular Fock state. There is no contribution
from $H_{0}G_{\overline{10}}$, $H_{1}G_{1}$ and
$H_{2}G_{\overline{10}}$ as they form an antisymmetric sea under the
exchange of two gluons which makes these wave functions
antisymmetric and therefore unacceptable for a bosonic system (gg).
Equating the sum of all these partial probabilities to value of
probabilities $\rho _{0002}$, $\rho _{2000}$, $\rho _{0200}$ taken
from table 3 for $\Sigma^{*0}$ gives the unknown parameter c as: 23c
= 0.00599579 , $c_{0002}$= 0.0002606865217 and other values can be
computed and written as  $c_{2000}$= 0.000559926087, $c_{0200}$=
0.000559926087. Similar decompositions can be done for other Fock
states as well i.e by taking the sea upto three gluons as shown in
tables below.
\setlength{\tabcolsep}{0.9em} %
{\renewcommand{\arraystretch}{1.3}%

\FloatBarrier
\begin{table}[H]
\caption{Computed probability ratios for various Fock states in spin
and color space. }
\begin{center}
\begin{tabular}{|c|c|c|c|c|c|}
\hline
Probability Ratio $\rightarrow$ & $\frac{\rho_{\frac{1}{2},1}}{\rho_{\frac{1}{2},2}}$ &$\frac{\rho_{\frac{3}{2},0}}{\rho_{\frac{3}{2},1}}$& $\frac{\rho_{\frac{3}{2},0}}{\rho_{\frac{3}{2},2}}$  & $\frac{\rho_{1,1}}{\rho_{8,8}}$ &$\frac{\rho_{1,1}}{\rho_{10,\overline{10}}}$ \\
States $\downarrow$   & & & &&\\\hline $|gg\rangle$&$\frac{1}{2}$ &2
&$\frac{1}{2}$ &$\frac{1}{2}$ &1\\\hline $|u\overline{u}g\rangle$ &
$\frac{1}{2}$&2 &$\frac{1}{2}$ &$\frac{1}{4}$ &1\\\hline
$|d\overline{d}g\rangle$ & $\frac{1}{2}$&2 &$\frac{1}{2}$
&$\frac{1}{4}$ &1\\\hline $|s\overline{s}g\rangle$ & $\frac{1}{2}$&2
&$\frac{1}{2}$ &$\frac{1}{4}$&1\\\hline $|u\overline{u}
d\overline{d}\rangle$ & $\frac{1}{2}$&2 &$\frac{1}{2}$
&$\frac{1}{4}$ &1\\\hline $|u\overline{u} s\overline{s}\rangle$ &
$\frac{1}{2}$&2 &$\frac{1}{2}$ &$\frac{1}{4}$ &1\\\hline
$|s\overline{s} d\overline{d}\rangle$ & $\frac{1}{2}$&2
&$\frac{1}{2}$ &$\frac{1}{4}$ &1\\\hline $|d\overline{d}
d\overline{d}\rangle$ &$\frac{1}{2}$ &2 &$\frac{1}{2}$
&$\frac{1}{2}$ &1\\\hline $|u\overline{u} u\overline{u}\rangle$
&$\frac{1}{2}$ &2 &$\frac{1}{2}$ &$\frac{1}{2}$ &1\\\hline
\end{tabular}
\end{center}
\end{table}
\FloatBarrier

\setlength{\tabcolsep}{0.5em} %
{\renewcommand{\arraystretch}{1.8}%
\FloatBarrier
\begin{table}[H]
\caption{Computed probability ratios for various Fock states in spin
and color space.}
\begin{center}
\begin{tabular}{|c|c|c|c|c|c|c|}
\hline Probability Ratio $\rightarrow$ &
$\frac{\rho_{\frac{1}{2},1}}{\rho_{\frac{1}{2},2}}$ &
$\frac{\rho_{\frac{3}{2},1}}{\rho_{\frac{3}{2},2}}$&
$\frac{\rho_{\frac{1}{2},1}}{\rho_{\frac{3}{2},1}}$
&$\frac{\rho_{\frac{3}{2},1}}{\rho_{\frac{3}{2},0}}$ &
$\frac{\rho_{1,1}}{\rho_{8,8}}$ &$\frac{\rho_{1,1}}{\rho_{10,\overline{10}}}$ \\
States $\downarrow$   &S,A&S,A &S,A &S,A&S,A&S,A \\\hline
$|d\overline{d} d\overline{d}g\rangle$
&1,$\frac{1}{2}$&$\frac{3}{2}$,$\frac{3}{4}$
&$\frac{1}{3}$,$\frac{1}{3}$&3
&$\frac{1}{8}$,$\frac{1}{8}$&$\frac{1}{2}$,$\frac{1}{2}$\\\hline
$|d\overline{d} s\overline{s}g\rangle$
&1,$\frac{1}{2}$&$\frac{3}{2}$,$\frac{3}{4}$
&$\frac{1}{3}$,$\frac{1}{3}$&3
&$\frac{1}{8}$,$\frac{1}{8}$&$\frac{1}{2}$,$\frac{1}{2}$\\\hline
$|u\overline{u} gg\rangle$
&1,$\frac{1}{2}$&$\frac{3}{2}$,$\frac{3}{4}$
&$\frac{1}{3}$,$\frac{1}{3}$&3
&$\frac{1}{8}$,$\frac{1}{8}$&$\frac{1}{2}$,$\frac{1}{2}$\\\hline
$|d\overline{d}gg\rangle$
&1,$\frac{1}{2}$&$\frac{3}{2}$,$\frac{3}{4}$
&$\frac{1}{3}$,$\frac{1}{3}$&3
&$\frac{1}{8}$,$\frac{1}{8}$&$\frac{1}{2}$,$\frac{1}{2}$\\\hline
$|s\overline{s} gg\rangle$
&1,$\frac{1}{2}$&$\frac{3}{2}$,$\frac{3}{4}$
&$\frac{1}{3}$,$\frac{1}{3}$&3
&$\frac{1}{8}$,$\frac{1}{8}$&$\frac{1}{2}$,$\frac{1}{2}$\\\hline
$|u\overline{u} u\overline{u}g\rangle$
&1,$\frac{1}{2}$&$\frac{3}{2}$,$\frac{3}{4}$
&$\frac{1}{3}$,$\frac{1}{3}$&3
&$\frac{1}{8}$,$\frac{1}{8}$&$\frac{1}{2}$,$\frac{1}{2}$\\\hline
$|u\overline{u} d\overline{d}g\rangle$(no symmetry condition)
&$\frac{3}{4}$&$\frac{9}{8}$ & -&$\frac{9}{2}$
&$\frac{1}{8}$&$\frac{1}{2}$\\\hline
\end{tabular}
\end{center}
\end{table}
\FloatBarrier

\subsection{Modifications in statistical model}
Statistical model and detailed balance principle is used in
combination to study the properties like magnetic moment and
$\overline{d}-\overline{u}$ asymmetry, vector and scalar dominancy
in the sea and strange quark importance to the magnetic moment of
decuplets. There are three modification in specific, that we have
focussed upon, and they are discussed below.

\medskip Model D assumes, that sea containing large number of gluons have relatively
smaller probabilities and hence their multiplicities have been
suppressed over the rest of valence particles with limited quarks.
In order to check the predicting power of statistical approach, we
have modified the relative probabilities by suppressing the
contribution of states coming from higher multiplicities. Here
relative probabilities are divided with their respective spin and
color multiplicities to achieve the suppression. This modification
is based on the phenomenological ground [37] stating, that the
higher the multiplicities, lower will be the associated
probabilities. The decompositions of Fock states with this new input
is shown as:

1. $|gq\overline{q}\rangle, |u\overline{u}d\overline{d}\rangle,
|u\overline{u}s\overline{s}\rangle,
|d\overline{d}s\overline{s}\rangle $ sea , symmetry consideration is
not needed.
\begin{eqnarray*}
\rho_{\frac{1}{2},1}[\rho_{1,1},\rho_{8,8},\rho_{10,\overline{10}}]=c(1,4,1)=d(\frac{1}{3},\frac{1}{48},\frac{1}{300})\\
\rho_{\frac{1}{2},2}[\rho_{1,1},\rho_{8,8},\rho_{10,\overline{10}}]=2c(1,4,1)=2d(\frac{1}{5},\frac{1}{80},\frac{1}{500})\\
\rho_{\frac{3}{2},0}[\rho_{1,1},\rho_{8,8},\rho_{10,\overline{10}}]=c(1,4,1)=d(\frac{1}{2},\frac{1}{32},\frac{1}{200})\\
\rho_{\frac{3}{2},1}[\rho_{8,8}]=4c=\frac{d}{96}\\
\rho_{\frac{3}{2},2}[\rho_{1,1},\rho_{8,8},\rho_{10,\overline{10}}]=2c(1,4,1)=2d(\frac{1}{10},\frac{1}{160},\frac{1}{1000})
\end{eqnarray*}
Summing and equating all the partial probabilities to
$\rho_{1001},\rho_{0101},\rho_{0011},\rho_{1100},\rho_{1010},\rho_{0110}$
we get values of d as: 0.01069737, 0.01069737, 0.044775695,
0.022387718, 0.022387718, 0.022387718 respectively. All the
calculations of relative probability $\rho_{p,q}$ for spin and color
spaces with the possibilities arising from gluons are shown in table
below. Similar numbers can be obtained for other Fock states as
well. Details of the calculations is given in Ref. [37].

\medskip Detailed balance principle is applied to put a limit on the number
of $s\overline{s}$ pairs in the sea (due to the fixed mass of
decuplets), in terms of constraint as $(1-C_{l})^{n-1}$ where
$C_{l}=\frac{2M_{s}}{M_{B}-2lM_{s}}$ ($M_{B}$=mass of baryon,
$M_{s}$=mass of strange quark, n is the total number of partons.
This kind of constraint is proven to be helpful to understand the
strange behavior of the sea in various decuplets. It has been
noticed in general that, the strange sea dominates over non-strange
sea quarks for strange baryon particles
($\Sigma^{*+},\Sigma^{*0},\Sigma^{*-},\Xi^{*0},\Xi^{*-},\Omega^{-}$)
as compared to non strange decuplet members
($\Delta^{++},\Delta^{+},\Delta^{0},\Delta^{-}$).

\medskip To appreciate the importance of sea with spin, modifications
in the model is done by choosing sea to be contributing through
scalar, vector or tensor coefficients. Here sea with spin 0,1,2 are
called scalar, vector and tensor sea respectively. These
coefficients are directly related to probabilities of quark-gluon
Fock states in spin color and flavor space. Here sea is found to be
dynamic for scalar and tensor part unlike baryon octets where tensor
appears to be less dominating due to quark spin flip processes [38].
In specific, magnetic moments get influenced by suppressing any of
the parameter/coefficient in the total wavefunction.

\medskip The importance of SU(3) symmetry and its breaking has been discussed for baryon
octets [38]. Due to the limited experimental information on
SU(3)symmetry breaking in decuplets, we have restricted ourselves to
analyse $\overline{d}-\overline{u}$ asymmetry contributing to the
magnetic moment of decuplets. These data may be useful for
experimentalist to investigate further.

Due to the variation in the individual contributions coming from sea
containing either $u\overline{u},d\overline{d}$ and $s\overline{s}$
pairs motivates us to study the $\overline{d}-\overline{u}$
asymmetry in the valence and sea quarks. The anti-quark flavor
asymmetry $\overline{d}-\overline{u}$ is calculated for all baryon
decuplets to justify their importance in the sea. The set of
experiments like SLAC in 1975 [39], E288 collaboration with
Drell-Yan experiment at Fermilab in 1981 [40], HERMES collaboration
at DESY [41-43] have been successful in measuring
$\overline{d}-\overline{u}$ asymmetry over a period of time. On
phenomenological grounds, meson cloud model and chiral quark model
have also been able to confirm the values observed experimentally.
Also, Zhang et al. have used principle of detailed balance to
calculate flavor asymmetry and predicted the value as 0.118 [43]
matching well with experimental value of 0.124 [42].

\section{Results and Discussion}
Magnetic moments of baryon decuplets are calculated in statistical
model in which the hadronic structure is considered to be consisting
of valence quarks and sea limited to few number of quarks and gluon
states multiconnected non perturbatively through gluons. The
calculation of magnetic moments are performed by two approaches i.e.
Model C and Model D. Model C aims at finding relative probabilities
of the Fock states in color, spin and flavor space whereas Model D
finds the probabilities of Fock states by suppressing the
contribution of states with higher multiplicities. The conclusions
of this approach are as follows:
\begin{itemize}
\item Magnetic moments of decuplets are shown in table 6. The
results are based on a combined approach of statistical model and
detailed balance principle. The experimental information on magnetic
moment of $\Delta^{++}, \Delta^{+}, \Omega^{-}$ [46-48] are very
well retrieved with our approach other moments are compared with the
theoretical data and found to be matching well within the error bar
10-20\%. C model is showing better match than the D model where
suppressed multiplicity are used to calculate relative
probabilities. Further modification in C model leads to the
following conditions. The modification is done in terms of
assumptions where sea can be strange or non-strange quark-gluon Fock
states.

\item
It can be concluded that probability of strange sea is more in
strange particles as compared to non strange ones. The reason being
that strange quarks contribute very less as compared to up and down
quarks which are lighter than heavy (strange) quarks. Here, the
importance of constraint applied in the procedure can be justified
as:

\begin{center}
\begin{tabular}{|c|c|}
  \hline
Fock state & Value of constraint $(1-C_{l})^{n-1}$ \\\hline
  uds $u\overline{u}s\overline{s}g$ & 0.271  \\
  uss $u\overline{u}s\overline{s}g$ & 0.317\\\hline
\end{tabular}
\end{center}
The data clearly shows the more importance of constraint on doubly
strange baryon than single strange baryon for one $s\overline{s}$
pair in sea.

\item Along with this, the effect of suppressing the contribution from scalar, vector
and tensor sea has also been estimated separately. Table 7 shows the
contributions from scalar-tensor and vector sea. Individual
contributions from scalar, vector and tensor sea was checked which
concluded that scalar-tensor sea gives a good match with the
observed values in our model. Vector sea contributions was
suppressed to check the contributions from the scalar-tensor sea and
similarly other contributions could be calculated. So, scalar-tensor
sea dominance is confirmed to produce results closer to experimental
data.

\item The $\overline{d}-\overline{u}$ asymmetry is calculated for the decuplet particles using principle of detailed
balance and is shown in table 7.

\item Besides this, the value magnetic moment ratio
$\frac{\mu(\Delta^{++})}{\mu(p)}= 1.97$ is in good agreement with
the prediction of simple quark model i.e.
$\frac{\mu(\Delta^{++})}{\mu(p)}= 2.0$ and experimentally the ratio
is predicted to be 1.62$\pm$0.18 . Interestingly, our value of
$\frac{\mu(\Omega^{-})}{\mu(\Lambda^{0})}$ = 2.98 also matches with
the predicted value by simple quark model equal to 3.

\item Our model has been able to produce excellent fit to the sum rules given by
Soon-Tae Hong [49] for baryon decuplet magnetic moments i.e
\begin{eqnarray*}
\mu_{\Sigma^{*0}}=\frac{1}{2}\mu_{\Sigma^{*+}}+\frac{1}{2}\mu_{\Sigma^{*-}}\\
\mu_{\Delta^{-}}+\mu_{\Delta^{++}}=\mu_{\Delta^{0}}+\mu_{\Delta^{+}}
\end{eqnarray*}
\end{itemize}

\setlength{\tabcolsep}{0.4em} %
{\renewcommand{\arraystretch}{1.5}%
\FloatBarrier
\begin{table}[H]
\caption{Comparison of computed magnetic moments (in terms of
$\mu_{N}$) of baryon decuplet with other models and experimental
data.}
\begin{center}
\begin{tabular}{|c|c|c|c|c|c|c|c|c|}
\hline
Particle &C Model&D Model&SQM &QCDQM[17]&$\chi$QM[44]&CQSM[22]&CBM[45]&Data \\
\hline
$\Delta^{++}$&5.50  &4.82&5.56& 5.689&5.30&4.85&4.52&4.52$\pm$0.50 [46]\\
$\Delta^{+}$& 2.69 & 2.36&2.73& 2.778&2.58&2.35&2.12&$2.7^{+1.0}_{-1.3}\pm1.5\pm3$[47]\\
$\Delta^{0}$ & -0.09 & -0.13&-0.09&-0.134&-0.13&-0.14&-0.29&-\\
$\Delta^{-}$& -2.88 & -2.52&-2.92&-3.045&-2.85&-2.63&-2.69&-\\
$\Sigma^{*+}$ & 3.05 & 2.93&3.09&2.933&2.88&2.47&2.63&-\\
$\Sigma^{*0}$ & 0.26 & 0.22&0.27&0.137&0.17&-0.02&0.08&-\\
$\Sigma^{*-}$ & -2.53 & -2.23&-2.56&-2.659&-2.55&-2.52&-2.48&-\\
$\Xi^{*0}$& 0.61 & 0.47&0.63 &0.424&0.47&0.09&0.44&-\\
$\Xi^{*-}$& -2.16 & -1.84&-2.31&-2.307&-2.25&-2.40&-2.27&-\\
$\Omega^{-}$& -1.82 &-1.67&-1.84&-1.970&-1.95&-2.29&-2.06&-2.02$\pm$0.05[48]\\
\hline
\end{tabular}
\end{center}
\end{table}
\FloatBarrier

\setlength{\tabcolsep}{0.3em} %
{\renewcommand{\arraystretch}{1.4}%
\FloatBarrier
\begin{table}[H]
\caption{Various modifications in Statistical Model }
\begin{center}
\begin{tabular}{|c|c|c|c|c|c|c|}
\hline
& \multicolumn{2}{p{3.5cm}|}%
{\centering C model}&&&&\\
\cline{2-3} \multicolumn{1}{|c|}{Particle}
& \multicolumn{1}{p{2.7cm}|}{\centering With strange sea \\
$(g\rightarrow u\overline{u},d\overline{d},s\overline{s}$)} &
\multicolumn{1}{p{2.8cm}|}{\centering Without $s\overline{s}$ sea
\\(only $g\rightarrow u\overline{u},d\overline{d}$)}
&\multicolumn{1}{c|}{$\overline{d}-\overline{u}$ asymmetry}
&\multicolumn{1}{c|}{Vector sea}
&\multicolumn{1}{p{2.0cm}|}{\centering Scalar-Tensor \\ sea}
&\multicolumn{1}{c|}{Data}
\\
\hline

$\Delta^{++}$& 5.50  & 5.51&0.4&4.07&5.50&4.52$\pm$0.50\\
$\Delta^{+}$& 2.69  &2.70  &0.12&2.003&2.70&$2.7^{+1.0}_{-1.3}\pm1.5\pm3$\\
$\Delta^{0}$ & -0.09  & -0.09&-0.12& -0.067&-0.096&-\\
$\Delta^{-}$&-2.88 &-2.89&-0.4 & -2.13&-2.88&-\\
$\Sigma^{*+}$ &3.05 & 3.01&0 & 2.26&3.05 &-\\
$\Sigma^{*0}$ &0.26  &0.26 &0 & 0.19&0.26&-\\
$\Sigma^{*-}$ &-2.53  & -2.51&-0.37 & -1.87&-2.53&-\\
$\Xi^{*0}$&0.61  &0.59&0.26 &0.45&0.61&-\\
$\Xi^{*-}$&-2.16 &-2.11&-0.26&-1.61&-2.17&-\\
$\Omega^{-}$&-1.82 &-1.71&0&-1.34&-1.82&-2.02$\pm$0.05\\
\hline
\end{tabular}
\end{center}
\end{table}
\FloatBarrier

\section*{Acknowledgement}
The authors gratefully acknowledge the financial support by the
Department of Science and Technology (SB/FTP/PS-037/2014), New
Delhi.

\bibliography{latex-sample}
\bibliographystyle{unsrt}
\begin{enumerate}
    \item  S. Chatrchyan et al. (CMS Collaboration), Phys. Rev. Lett. 108,
252002 (2012); S. Chatrchyan et al. (CMS Collaboration), JHEP 07,
163 (2013).

\item S. Coleman and S. L. Glashow, Phys. Rev. Lett. 6, 423(1961).

\item K. A. Olive et al. (Particle Data Group), Chin. Phys. C, 38, 090001
(2014).

\item A. Bosshard, C. Amsler, M. Doebeli, M. Doser, M. Schaad, J. Riedlberger, P. Truoel and
J. A. Bistirlich et al., Phys. Rev. D 44, 1962 (1991).

\item H. T. Diehl, S. Teige, G. B. Thomson, Y. Zou, C. James, K. B. Luk, R. Rameika and
P. M. Ho et al., Phys. Rev. Lett. 67, 804 (1991).

\item N. B. Wallace, P.M. Border, D. P. Ciampa, G. Guglielmo, K. J. Heller, D. M. Woods, K.
A. Johns and Y. T. Gao et al., Phys. Rev. Lett. 74, 3732 (1995).

\item M.Kotulla et al. Phys. Rev. Lett. 89, 272001 (2002).

\item  I.V. Gorelov (CDF Collaboration) arXiv:hep-ex/0701056,(2007).

\item J. Ashman, et al.,(EMC Collaboration) Phys. Lett. B 206 (1988) 364.
\item B. Adeva et al. (SMC Collaboration), Phys. Rev. D 58, 112001 (1998).

\item X. Song, V. Gupta, Phys. Rev. D. 49 , 2211-2220, (1994).

\item H. Dahiya and M. Gupta, Phys. Rev. D 67, 114015 (2003).

\item M.D. Slaughter, Phys. Rev. C 82, 015208 (2010); M.D. Slaughter, Phys. Rev. D 84, 071303
(2011);J. Linde, T. Ohlsson, and H. Snellman, Phys. Rev. D 57, 452
(1998); J. Linde, T. Ohlsson, and H. Snellman, Phys. Rev. D 57, 5916
(1998).

\item I.S. Sogami and Oh'Yamaguchi, Phys. Rev. Lett. 54
(1985) 2295.

\item B. S. Bains and R. C. Verma, Phys. Rev. D 66, 114008 (2002).

\item F. Schlumpf, Phys. Rev. D 48, 4478 (1993); G. Ramalho, K. Tsushima, and F. Gross, Phys.
Rev. D 80, 033004 (2009).

\item P.Ha, Phys. Rev. D 58, 113003 (1998); C.S. An, Q.B. Li, D.O. Riska, and B.S. Zou, Phys.
Rev. C 74, 055205 (2006).

\item B.S. Bains and R.C. Verma, Phys. Rev D 66, 114008 (2002); R. Dhir and R.C. Verma, Eur.
Phys. J. A 42, 243 (2009).

\item T.M. Aliev, A. Ozpineci, and M. Savci, Phys. Rev. D 62, 053012 (2000).

\item F.X. Lee, Phys. Rev. D 57, 1801 (1998); S.L. Zhu, W.Y.P. Hwang, and
Z.S.P. Yang, Phys. Rev. D 57, 1527 (1998); A. Iqubal, M. Dey, and J.
Dey, Phys. Lett. B 477, 125 (2000).

\item B. Schwesinger and H. Weigel,
Nucl. Phys. A 540, 461 (1992); Y. Oh, Phys. Rev. D 75, 074002
(2007).

\item T. Ledwig, A. Silva, and M. Vanderhaeghen, Phys. Rev D 79 094025 (2009).
;H-C. Kim, M. Praszalowicz, and K. Goeke, Phys. Rev. D 57, 2859
(1998); G.S. Yang, H-C. Kim, M. Praszalowicz, and K. Goeke, Phys.
Rev. D 70, 114002 (2004). ;R. Flores-Mendieta, Phys. Rev. D 80,
094014 (2009); L.S. Geng, J.M. Camalich, and M.J.V. Vacas, Phys. Rev
D 80, 034027 (2009).

\item R. Flores-Mendieta, Phys. Rev. D 80, 094014 (2009); L.S. Geng, J.M.
Camalich, and M.J.V. Vacas, Phys. Rev D 80, 034027 (2009).

 \item S. Boinepalli, D.B. Leinweber, P.J. Moran,
A.G.Williams, J.M. Zanotti, and J.B. Zhang, Phys. Rev. D 80, 054505
(2009); C. Aubin, K. Orginos, V. Pascalutsa, and M. Vanderhaeghen,
Phys. Rev. D 79, 051502 (2009); P.E. Shanahan et al. (CSSMand
QCDSF/UKQCD Collaborations), Phys. Rev. D 89, 074511 (2014).

\item F. X. Lee, R. Kelly, L. Zhou and W. Wilcox, Phys. Lett. B 627, 71
(2005).

\item  A.O. Bazarko et al.,  Z.Phys. C65 (1995) 189-198.

\item A. Acha, et al., HAPPEX Collaboration, Phys. Rev. Lett. 98 (2007)
032301.

\item S. Baunack, et al., Phys. Rev. Lett. 102 (2009) 151803.

\item D. Androic, et al., G0 Collaboration, Phys. Rev. Lett. 104 (2010)
012001.

\item R. Bijker, et al., Phys. Rev. C 85 (2012) 035204.

\item J. F. Donoghue and E. Golowich, Phys. Rev. D 15, 3421
(1977);He Hanxin, Zhang Xizhen, and Zhuo Yizhong, Chinese Phys. 4,
359 (1984).

\item E. Golowich, E. Haqq, and G. Karl, Phys. Rev. D 2, 160
(1983); F. E. Close and Z. Li, Phys. Rev. D 42, 2194 {1990}; F. E.
Close, Rep. Prog. Phys. 51, 833 (1988); Z. Li, Phys. Rev. D 44, 2841
(1991).

\item Y.J. Zhang, W.Z. Deng, B.Q. Ma, Phys. Lett. B 523 (2002) 260; Y.J. Zhang, B.Q. Ma, L. Yang, Int. J. Mod. Phys. A 18 (2003)
1465–1468.

\item Y.J. Zhang, B. Zhang, B.Q. Ma, Phys. Rev. D 65 (2002)
114005.

\item M.Batra, A.Upadhyay, Int.J.Mod.Phys. A28 (2013) 1350062.

\item J. P. Singh, A. Upadhyay, J. Phys. G, Nucl. Part. Phys. 30 (2004) 881–894.

\item A. Upadhyay, M. Batra, The European Physical Journal A,
(2013).

\item M. Batra et al. J.Phys.Conf.Ser. 481 (2014) 012024.

\item S.Stein et al., Phys. Rev D 12,1884(1975).

\item A. S. Ito et al.[E288 Collaboration], Phys. Rev. D
23,604(1981).

\item E. A. Hawker et al.,Phys. Rev. Lett. 80,3715(1998);J. C. Peng
et al.,Phys. Rev. D 58,092004(1998).

\item R. S. Towell et al.,Phys. Rev. D
64,052004(2001).

\item Y.J. Zhang, B. Zhang, B.Q Ma, Phys. Lett. B524, 260 (2001).

\item J. Linde, T. Ohlsson, and H. Snellman, Phys. Rev. D 57, 452 (1998); J. Linde, T. Ohlsson,
and H. Snellman, Phys. Rev. D 57, 5916 (1998).

\item S-T. Hong, Phys. Rev. D 76, 094029 (2007).

\item A.Bossard et al., Phys. Rev. D 44 (1991) 1962.

\item M. Kotulla et al. Phys. Rev. Lett. 89, 272001 (2002).

\item K.A. Olive et al. (Particle Data Group), Chin. Phys. C, 38, 090001 (2014).

\item S.T. Hong, G. E. Brown, Nuclear Physics A580 (1994) 408-418.

\end{enumerate}

\end{document}